\documentclass[aps,secnumarabic,showpacs,nobibnotes,amssymb,prd]{revtex4}
\begin{document}
\title{\bf Derivation of Einstein's Equation from a New Type of four Dimensional Superstring}
\medskip
\author{B. B. Deo and L. Maharana\footnote{E-Mail: lmaharan@iopb.res.in}}
\affiliation{ Physics Department, Utkal University, Bhubaneswar-751004, India.}
\begin{abstract}
A new type of superstring in four dimension is proposed which has the central charge 26.
The Neveu Schwarz and the Ramond vacua are both tachyonic. The self energy of the 
scalar tachyon cancels from the contribution of the fermionic loop of the Ramond sector. 
The NS tachyonic vacuum is used to construct a massless graviton. Coulombic 
vector excitations of zero mass, referred as a `Newtonian' graviton are also shown to exist.
The propagators are explicitly evaluated. Following the method of Weinberg, we
deduce the Einstein's field equation of general relativity.
\end{abstract}
\maketitle
The string theory has come to prominence due to appearance of the graviton in the 
mass spectrum of Nambu-Goto string. The researches on this line has been to develop 
gravity from a ten dimensional superstring by writing the group invariant but 
nonrenormalisable supergravity Lagrangian and studying the phenomenology 
arising therefrom at low energies. Recently one of us ~\cite{deo} has proposed a new 
type of superstring having central charge 26 and two tachyons whose self energies 
cancel. The salient features  are given below for completeness.

There is an effective dimensional reduction by adding 11 fermions which are the Lorentz
vectors in SO(3,1) bosonic sector, $\psi^{\mu,j}$, j=1,..,6 ; $\phi^{\mu,k}$, k=7,..,11 
in the world sheet ($\tau, \sigma$), to the 4 bosonic co-ordinates $X^{\mu}(\sigma,\tau)$. 
The action is~\cite{deo}
\begin{equation}
S= -\frac{1}{2\pi}\int d^2\sigma\left [ \partial_{\alpha}X^{\mu}\partial^{\alpha}X_{\mu}
-i \bar{\psi}^{\mu,j}\rho^{\alpha}\partial_{\alpha}  \psi_{\mu,j}
+ i \bar{\phi}^{\mu,k}\rho^{\alpha}\partial_{\alpha}  \phi_{\mu,k}\right ].\label{a}\\
\end{equation}
The upper indices (j,k) refer to rows and the lower indices to columns. Here
$\psi^{\mu,j }=\psi^{(+)\mu,j} +\psi^{(-)\mu,j}$ but \\$\phi^{\mu,k} = \phi^{(+)\mu,k} 
- \phi^{(-)\mu,k}$ where $+,-$ refer to the positive and the negative parts of the Majorana 
massless fermions. There are in all 13 metric ghosts, two bosonic and eleven fermionic.
In the Lorentz metric, the fermionic sector light cone action is\\
\[ S_{l.c.}=\frac{i}{2\pi}\;\int d^2\sigma \sum_{\mu =0,3}\left( \bar{\psi}^{\mu j}\rho^
{\alpha}\partial_{\alpha}\psi_{\mu j} - \bar{\phi}^{\mu k}\rho^
{\alpha}\partial_{\alpha}\phi_{\mu k}\right) \]\\ 
contributes eleven to the central charge like the super conformal Faddeev-Popov ghosts.
They will be eliminated from the Fock space by the susidiary conditions. $\rho$-matrices
are given in ~\cite{deo} and ~\cite{gr} along with other objects.

The action (\ref{a}) is invariant under the transformations
\begin{eqnarray}
\delta X^{\mu} =\bar{\epsilon}\;(\;e^j\;\psi^{\mu}_j - e^k\;\phi^{\mu}_k),\\
\delta\psi^{\mu,j}= - i\;e^j\;\rho^{\alpha}\;\partial_{\alpha}X^{\mu}\;\epsilon,\\
\delta\phi^{\mu,k}= ie^k\;\rho^{\alpha}\partial_{\alpha}X^{\mu}\;\epsilon,
\end{eqnarray}
with $\epsilon$ as the constant anticommutating spinor. ${e}^j$ and
${e}^k$ are the two unit component c-number row vectors with the property that
${e}^j{e}_{j'}=\delta^j_{j'}$ and  ${e}^k{e}_{k'}=\delta^k_{k'}$ and it 
follows that 
${e}^j{e}_{j}$=6 and ${e}^k{e}_{k}$=5.

The commutator of two supersymmetric transformation is a translation
\begin{equation}
[ \delta_1 ,\delta_2 ] X^{\mu} = a^{\alpha}\partial_{\alpha}X^{\mu},
\end{equation}
where $a^{\alpha}=2i\bar{\epsilon}_1\rho^{\alpha}\epsilon_2$. 
Similar are the results for the spinors $\psi^{\mu,j}= e^j\Psi^{\mu}$ and
$\phi^{\mu,k}= e^k\Psi^{\mu}$ where
\begin{equation}
\Psi^{\mu}=e^j\;\psi^{\mu}_j - e^k\;\phi^{\mu}_k.
\end{equation}
It is easy to verify that
$\delta X^{\mu} =\bar{\epsilon}\Psi^{\mu}$, $\delta\Psi^{\mu}=-i\rho^{\alpha}
\partial_{\alpha}X^{\mu}\epsilon $ and
$[ \delta_1 ,\delta_2 ] \Psi^{\mu} = a^{\alpha}\partial_{\alpha}\Psi^{\mu}$.

Thus $\Psi^{\mu}$ is the supersymmetric partner of $X^{\mu}$. Introducing another 
supersymmetric pair, the zweibein $e_{\alpha}(\sigma,\tau)$ and the `gravitino' 
$\chi_{\alpha}=\nabla_{\alpha}\;\epsilon$, the local 2d supersymmetric action is~\cite{gr}
\begin{equation}
S= -\frac{1}{2\pi}\int d^2\sigma ~e~\left [ h^{\alpha\beta}\partial_{\alpha}X^{\mu }
\partial_{\beta}X_{\mu } -i\bar \Psi^{\mu}\rho^{\alpha}\nabla_{\alpha}
\Psi_{\mu}+ 2\bar{\chi}_{\alpha}\rho^{\beta}\rho^{\alpha}\Psi^{\mu}
\partial_{\beta}X_{\mu}+\frac{1}{2}
\bar{\Psi }^{\mu}\Psi_{\mu}\bar{\chi}_{\beta} \rho^{\beta}\rho^{\alpha}\chi_{\alpha}.
\right ]
\label{s1}.
\end{equation}
The variation of this equation with respect to $\chi^{\alpha}$ and $e^{\alpha}_b$ leads to
the equations for current and energy momentum tensors. In the gauge where the gravitino 
$\chi_{\alpha}$  wii be zero, and $e^a_{\alpha} = \delta^a_{\alpha}$,
\begin{equation}
J_{\alpha}= \frac{\pi}{2e}\frac{\delta S}{\delta \bar\chi^{\alpha}}=\frac{1}{2}\rho^{\beta}
\rho_{\alpha}\bar{\Psi}^{\mu}
\partial_{\beta}X_{\mu} - \frac{1}{2}\bar\Psi^{\mu}\Psi_{\mu}\rho_{\alpha}\rho^{\beta}
\chi_{\beta} = 0,
\end{equation}
\begin{equation}
T_{\alpha\beta}=\partial_{\alpha}X^{\mu }
\partial_{\beta}X_{\mu }- \frac{i}{2}\bar{\Psi}^{\mu}\rho_{(\alpha}\partial_{\beta )}
\Psi_{\mu}-(trace)=0.
\end{equation}
In the light cone basis $\tilde {\psi}=(\psi^+,\psi^-)$ and $\tilde {\phi}=(\phi^+,\phi^-)$,
the above equations translate to
\begin{equation}
J_{\pm}=\partial_{\pm}X_{\mu}\Psi^{\mu}_{\pm}=0\label{J},
\end{equation}
and
\begin{equation}
T_{\pm\pm}=
\partial_{\pm}X^{\mu}\;\partial_{\pm}X_{\mu}+\frac{i}{ 2}\psi^{\mu,j }_{\pm}\;\partial_{\pm}
 \psi_{\pm\mu,j}- \frac{i}{2}\phi_{\pm}^{\mu k}\;\partial_{\pm}\phi_{\pm\mu,k}=0\label{t}
\end{equation}
where ~$\partial_{\pm}=\frac{1}{2}(\partial_{\tau} \pm \partial_{\sigma})$. The component 
constraints are
\begin{eqnarray}
\partial_{\pm}X_{\mu}\;\psi_{\pm}^{\mu,j} = \partial_{\pm}X_{\mu}\;
e^j\;\Psi^{\mu}_{\pm }=0,~~~~~~~~j=1,2...6.\\
\partial_{\pm}X_{\mu}\;\phi_{\pm}^{\mu,k} = \partial_{\pm}X_{\mu}\;e^k\;\Psi^{\mu}
_{\pm }=0,~~~~~~~~k=7,...11.
\end{eqnarray}
These eleven constraints are enough to eliminate the extra eleven Lorentz metric 
fermionic ghosts from physical spectrum. However, there will be one current constraint, 
equation (\ref{J}), from which the eleven follow. So there will be eleven subsidiary 
current generators combinable to one and eleven pairs of $(\beta^j,~\gamma^j)$, 
$(\beta^k,~\gamma^k)$ ghosts combinable to one pair $(\beta,~\gamma)$ for the construction of
nilpotent BRST charge. 

We write the action $S_F^{l.c.}$ in terms of light cone superpartner
$\Psi^{\pm}$. The Hilbert space is supplemented by a space where
particles obey bose statistics and the light cone fermions are
Grassmanians
$\theta, \bar{\theta}$. In particular,
$\bar{\Psi}^+(\bar{\theta})=\frac{1}{2}\bar{\theta} \bar{\gamma}$ and
$\rho^{\alpha} \bar{\Psi}^-(\theta)=\frac{1}{2}\theta\beta^{\alpha}$.
Integrating over $\theta\bar\theta$, we obtain
$S_F^{l.c.}=-\frac{1}{2\pi}\int d^2\sigma
\bar{\gamma}\partial_{\alpha}\beta^{\alpha}$. Standard text book results follow.
For the two groups of fermions, the group invariant constraints are
\begin{equation}
\partial_{\pm}X_{\mu}\;e^k\;\phi^{\mu}_{\pm k} = \partial_{\pm}X_{\mu}\;
e^j\;\psi^{\mu}_{\pm j}=0.\label{5}
\end{equation}

In the covariant formulation, the total number of physical degrees of freedom is the 
total number forty four minus the total number of constraints. Due to the above four 
constraints (~\ref{5}), there are 40 physical space time fermionic modes in the theory.\\ 

SO(3,1) has Dirac spinorial representation denoted by $\theta_{j,\delta}$ and 
$\theta_{k,\delta} ,where~~\delta=1,2,3,4$. So we can construct a genuine space time spinor 
with four components
\begin{equation}
\theta_{\delta}=\sum^6_{j=1}e^j\theta_{j\delta} -
\sum^{11}_{k=7}e^k\theta_{k\delta}.
\end{equation}

The Green Schwarz action~\cite{gr} exhibiting local four dimensional N=1 supersymmetry is
\begin{equation}
S=\frac{1}{2\pi}\int d^2\sigma \left ( \sqrt{g}g^{\alpha\beta}\Pi_{\alpha}\cdot\Pi_{\beta}
+2i\epsilon^{\alpha\beta}\partial_{\alpha}X^{\mu}\bar{\theta}\Gamma_{\mu}
\partial_{\beta}\theta \right ),
\end{equation}
where $\Gamma_{\mu}$ are the Dirac gamma matrices and
\begin{equation}
\Pi^{\mu}_{\alpha}=\partial_{\alpha}X^{\mu}- i\bar{\theta}\Gamma^{\mu}\partial_
{\alpha}\theta.
\end{equation}
It is difficult to quantise this action, so we fall back on the Neveu-Schwarz~\cite{ne} 
and the Ramond~\cite{ra} formulations with the G.S.O~\cite{gl} projection. 
The GSO operator is to project out the odd number of fermionic modes from the Hilbert
space and is defined as
\begin{equation}
G= \frac{1}{2}\left ( 1 +(-)^F\right ).
\end{equation}
where F is the fermion number. The forty fermionic modes can be placed in five identical
groups, each group containing eight of them. The total partition function is that of a
group of eight, raised to the power of five. It has been shown by Seiberg and Witten
~\cite{se} that the partition function of eight fermions with Neveu-Schwarz and Ramond
boundary conditions, vanish due to the famous Jacobi equality among the 
$\Theta$-functions. Thus the total product partition functions of the string states 
vanish. This is also the condition for a local supersymmetry.\\

The commutators and anticommutators between fields follow from the action of 
equation(\ref{a}). The fields can be quantised in the usual way~\cite{gr}. Let 
the bosonic quanta be denoted by $\alpha^{\mu}_m$, the ($b^{\mu}_{r,j},\; 
b^{\prime\mu}_{r,k}$) with r half integral be the quanta of ($\psi^{\mu}_j $, 
$\phi^{\mu}_j$) satisfying NS boundary condition and ($d^{\mu}_{m,j}$ ,
$\; d^{\prime\mu}_{m,k}$) with m integral, satisfying Ramond boundary condition. 
The nonvanishing commutation and anticommutation relations are
\begin{equation}
[\alpha_m^{\mu}, \alpha_n^{\nu}]=m\delta_{m,-n}g^{\mu\nu},
\end{equation}
\begin{eqnarray}
\{ b^{\mu ,j}_r , b^{\nu,j'}_s\}&=&g^{\mu\nu}\delta^{j,j'}\delta_{r,-s} ,\nonumber\\
\{ b^{'\mu ,k}_r , b^{'\nu,k'}_s\}&=& -g^{\mu\nu}\delta^{k,k'}\delta_{r,-s},\\
\{ d^{\mu ,j}_m , d^{\nu,j'}_n \}&=&g^{\mu\nu}\delta^{j,j'}\delta_{m,-n}, \nonumber\\
\{ d^{'\mu ,k}_m, d^{'\nu,k'}_n \}&=&-g^{\mu\nu}\delta^{k,k'}\delta_{m,-n}.
\end{eqnarray}
The phase of the creation operator is such that for ~~r,~m $ >$ 0, ~~$b_{-r}^{\prime}
=-b_r^{\prime\dag}$~~ and ~~$d^{\prime}_{-m}=-d_m^{\dag}$~~
It is necessary to identify the relations ~~$b_r^{\mu,j}~=~e^jB_r^{\mu},~~
b_r^{\prime\mu,k} = e^k B_r^{\mu}$, such that~~ 
$B_r^{\mu}=e^j b_{r,j}^{\mu} - e^k b_{r,k}^{\prime\mu}$.~ Similarly for the ~d,
~$d^{\prime}$; ~$D_m^{\mu}= e^j d_{m,j}^{\mu}- e^k d_{m,k}^{\prime\mu}$.~~
The Virasoro  generators
\begin{eqnarray}
L_m&=&\frac{1}{\pi}\int_{-\pi}^{\pi}d\sigma e^{im\sigma}T_{++}\nonumber\\
 &= &\frac{1}{2}\sum^{\infty}_{-\infty}:\alpha_{-n}\cdot\alpha_{m+n}: +\frac{1}{2}
\sum_{r\in z+\frac{1}{2}}(r+\frac{1}{2}m): (b_{-r} \cdot b_{m+r} - b_{-r}' \cdot b_{m+r}'):~~~~~~~~~NS
\nonumber\\
&=&\frac{1}{2}\sum^{\infty}_{-\infty}:\alpha_{-n}\cdot\alpha_{m+n}: +\frac{1}{2}
\sum^{\infty}_{n=-\infty}(n+\frac{1}{2}m): (d_{-n} \cdot d_{m+n} - d_{-n}'
\cdot d_{m+n}'):~, ~~~~~~~R\\
G_r &=&\frac{\sqrt{2}}{\pi}\int_{-\pi}^{\pi}d\sigma e^{ir\sigma}J_{+}=
\sum_{n=-\infty}^{\infty}\alpha_{-n}\cdot B_{n+r}, ~~~~~~~~~~~~~~~~~~~~~~~~~~~~~~ NS\\
F_m &=&\sum_{-\infty}^{\infty} \alpha_{-n}\cdot D_{n+m}.~~~~~~~~~ ~~~~~~~~~~~~~~~~~~~~~~~~~~~~~~~~~~~~~~~~~~~~~~~~ R
\end{eqnarray}
satisfy the super Virasoro algebra~\cite{v}
\begin{eqnarray}
\left [L_m , L_n\right ] & = &(m-n)L_{m+n} +\frac{C}{12}(m^3-m)\delta_{m,-n},\\
\left [L_m , G_r\right ] & = &(\frac{1}{2}m-r)G_{m+r}, ~~~~~~~~~~~~~~~~~~~~~~~~~~~~~~NS\\
\{G_r , G_s\} & =& 2L_{s+r} +\frac{C}{3}(r^2-\frac{1}{4})\delta_{r,-s},\\
\left [L_m , F_n\right ] & = & (\frac{1}{2}m-n)F_{m+n},~~~~~~~~~~~~~~~~~~~~~~~~~~~~~R\\
\{F_m, F_n\} & = & 2L_{m+n} +\frac{C}{3}(m^2-1)\delta_{m,-n},\;\;\;\; m\neq 0.
\end{eqnarray}
Since, from equation~(\ref{t}),~~$<T_{\pm\pm}(z)T_{\pm\pm}(\omega)>~=~\frac{26}{2}(z-\omega)^{-4}$ + ......, the central charge C=26. It is worth while to note that
$<T_{\pm\pm}^{l.c.}(z)T_{\pm\pm}^{l.c.}(\omega)>~=~\frac{11}{2}(z-\omega)^{-4}$ 
with central charge 11. The terms, containing the central charge  C=26, are the anomaly 
terms due to the normal ordering~\cite{gr}. As is well known, they are cancelled by the 
contribution from the conformal ghosts (b,c).   

This is also known that all anomalies will cancel if the normal ordering constant of 
$L_o$ is equal to one.
We define the physical states as satisfying
\begin{eqnarray}
(L_o-1)|\phi>&=&0,~~~L_m|\phi>=0,~~~G_r|\phi>=0~~~ for~~~~ r,m>0,\;\;\;\; :NS\;\;\;Bosonic\\
 L_m|\psi>&=&F_m|\psi>=0\;\;\;\;\;\;\;\;\;\;\;\;\;\;\;\;\;\;\;\;\;\;\;\;\;\;\;\;\;
for\;\;\;\;\; m>0,\;\;\;\;\;\;\;\; ~~:R\;\;\;\;Fermionic\\
(L_o-1)|\psi>_{\alpha}&=&(F_o^2-1)|\psi>_{\alpha}=0.
\end{eqnarray}
So we have
\begin{equation}
(F_o +1)|\psi_+>_{\alpha}=0\;\;\;\; and\;\;\; (F_o-1)|\psi_->_{\alpha}=0
\end{equation}
These conditions shall make the string model ghost free. 

It can be seen in a simple way. Applying $L_o$ condition
the state  $\alpha_{-1}^{\mu}|0,k>$ is massless and the $L_1$ constraint gives the
Lorentz condition $k^{\mu}|0,k>=0$ implying a transverse photon and
Gupta Bleuler impose that $\alpha_{-1}^0|\phi>=0$. Applying $L_2,\;L_3\; ....$, constraints,
one obtains $\alpha_{m}^0|\phi>=0$. Further, since $[\alpha_{-1}^0 , G_{r+1}]|\phi>=0$,$\;\;
B_r^0|\phi>=0$ and $b^0_{rj}|\phi>=e_{j}B^0_r|\phi>=0$; $b^{\prime 0}_{rk}|\phi>=
e_{k}B^0_r|\phi>=0$. All the time components are eliminated from
Fock space.

From fourier transforms and defination, the eleven subsidiary physical state conditions are
\begin{eqnarray}
G^j_r|\phi>&=&~~e^j G_r|\phi>=0,~~~G_r^k|\phi>=e^k G_r|\phi> = 0,~~~ 
for~~~~ r~~>~~0\label{gjr},\;\;\;\; NS\;\;\;\\
F^j_m|\phi>&=&~~e^j F_m|\phi>=0,~~~F_m^k|\phi>=e^k F_m|\phi> = 0,~~~ 
for~~~~ m~~>~~0.\label{fjm}\;\;\;\; R\;\;\;
\end{eqnarray}
We now proceed to write the nilpotent BRST charge. The part which comes from the usual 
conformal Lie algebra technique is
\begin{equation}
(Q_1)^{NS,R} = \sum (L_{-m}c_m)^{NS,R} -\frac{1}{2}\sum (m-n) : c_{-m}c_{-n}b_{m+n}:~ -
~a~ c_0 ;~~~~~Q^2_1=0 ~~~for ~a=1.
\end{equation}
The eleven pairs of commuting ghost quanta ($\beta_r,\gamma_r$) of ghost fields
($\beta(\tau),\gamma(\tau)$), satisfying $d\gamma = d\beta = 0$, needed for subsidiary current
generator operator conditions (\ref{gjr}) and (\ref{fjm}), are related as
\begin{eqnarray}
\beta^j_r& =&~~e^j \beta_r,~~~\beta_r^{'k} = e^k \beta_r, ~~~\beta_r = 
e^j \beta_{r,j}- e^k \beta'_{r,k}\label{b},\\
\gamma^j_r &=&~~e^j \gamma_r,~~~\gamma_r^{'k} = e^k \gamma'_r, ~~~\gamma_r = 
e^j \gamma_{r,j}- e^k \gamma'_{r,k}.~~~~(NS)
\end{eqnarray}
There are identical ones for the Ramond sector with half integral ~`r'~ replaced by integral~`m'~.
The light cone ghost quanta satisfy
~~$\{ b_r^{+,j},  b_s^{-,l}\}=-\delta_{r,-s}\delta^{j,l}$ ~~and
~~$\{ b_r^{'+,k},  b_s^{'-,m}\}=-\delta_{r,-s}\delta^{k,m}$, where as
~$[\gamma_s^j, \beta_r^{j'}]=\delta_{r,-s}\delta^{j,j'}$ and  
~$[ \gamma_s^k, \beta_r^{k'}]=\delta_{r,-s}\delta^{k,k'}$ and it follows that 
~$[ \gamma_s, \beta_r ]=\delta_{r,-s}$. 
All that we want to know is the conformal dimensions, $\gamma$ with ~$-\frac{1}{2}$,~ and
~$\beta$~ with ~$\frac{3}{2}$,
\begin{equation}
[~L_m, \gamma_r^{(j,k)}~]=-(\frac{3}{2}m +n)\gamma^{(j,k)}_{r+m},~~~~~~~
[~L_m, \beta_r^{(j,k)}~]=(\frac{1}{2}m -n)\beta^{(j,k)}_{r+m}.
\end{equation}
Using the Graded Lie algebra, we get the 
additional BRST charge. 
\begin{eqnarray}
Q_{NS}'&=&\sum G_{-r}\gamma_r -\sum\gamma_{-r}\gamma_{-s}b_{r+s},\nonumber\\
Q_{R}'&=&\sum F_{-m}\gamma_m -\sum\gamma_{-m}\gamma_{-n}b_{n+m}.
\end{eqnarray}
It is to be noted that the products \[ G_{-r}\gamma_{-r}=G^j_{-r}\gamma_{-r,j}-
G^k_{-r}\gamma_{-r,k},\] so that all the eleven pairs of ghosts are present in the charge.
As constructed, the BRST charge
\begin{equation}
Q_{BRST}=Q_1 + Q',
\end{equation}
is such that ~~$Q_{BRST}^2=0$~~in both NS and R sector~\cite{deo}. In
proving $\{Q',Q'\} + 2\{Q_1,Q'\}=0$, we have used the fourier
transforms, wave equations and integration by parts to show that
$\sum\sum
r^2\gamma_r\gamma_s\delta_{r,-s}=\sum_r\sum_s\gamma_r\gamma_s\delta_{r,-s}=0$.
The theory is unitary and ghost free.

The mass spectrum is
\begin{eqnarray}
NS:~~~~~~~\alpha^{\prime}M^2 &=& -1,-\frac{1}{2},0,\frac{1}{2},1,\frac{3}{2}....,\nonumber\\
R:~~~~~~~~\alpha^{\prime}M^2 &=& -1, 0, 1, 2, 3....~.
\end{eqnarray}
The GSO projection eliminates the half integral masses. The scalar bosonic vacuum 
energy $<0|(L^{NS}_o -1)^{-1}|0>$ is cancelled by the fermionic energy 
$-<0|(F_o -1)^{-1}(F_o +1)^{-1}|0>$, the negative sign arising due to the
normal ordering of the fermions. In both the sectors, we have the Regge trajectories
$\alpha^{\prime}M^2= -1, 0, 1, 2, 3....$ .\\

Thus, satisfying ourselves that we have an anomaly free, ghost free and harmless but useful tachyons, we attempt to tackle the problem 
of gravitational field theory from the above string theory.
Let us construct tensors like $b^{\mu}_{-\frac{1}{2} i}b^{\nu}_{-\frac{1}{2} j}$. For
simplicity, we drop the suffix $-\frac{1}{2}$. Consider the string state
\begin{equation}
{\bf a}^{\dag\mu\nu}(p)=\sum_{i,j} C^{ij} (b^{\mu}_i b^{\nu}_j + b^{\nu}_ib^{\mu}_j
-2\eta^{\mu\nu}b^{\lambda}_i b^{\lambda}_j)|0,p>.
\end{equation}
This is symmetric and traceless. Further $L_o$ will be taken as the free 
Hamiltonian $H_o$ in the interaction representation. Operating on this state, one gets~
$L_0{\bf a}^{\dag\mu\nu}(p)=0$. So this is massless. Further ~
$p_{\mu}{\bf a}^{\dag\mu\nu} = p_{\nu}{\bf a}^{\dag\mu\nu}=0$ ~if~ $C^{ij}=C^{ji}$.
If $C^{ij}$ is symmetric, a little algebra shows that 
\begin{equation}
G_{\frac{1}{2}}{\bf a}^{\dag\mu\nu}(p)= [ G_{\frac{1}{2}}, {\bf a}^{\dag\mu\nu}(p) ]=0.
\end{equation}
Thus ${\bf a}^{\dag\mu\nu}(p)$ satisfy all the physical state conditions due to 
Virasoro algebraic relations. This is the graviton. The commutator is
\begin{equation}
[{\bf a}^{\mu\nu}(p), {\bf a}^{\dag\lambda\sigma}(q)] = 2\pi |C|^2 \delta^4(p-q).\label{c}
\end{equation}
To switch over to the quantum field theory, we define the gravitational field by
space time fourier transform of ${\bf a}^{\mu\nu}(p)$ 
\begin{equation}
{\bf a}^{\mu\nu}(x)=\frac{1}{(2\pi)^3}\int \frac{d^3p}{\sqrt{2p^o}}\left [
{\bf a}^{\mu\nu}(p)\;e^{ipx} +{\bf a}^{\dag\mu\nu}(p)\;e^{-ipx}\right ],
\end{equation}
with the commutator
\begin{equation}
\;\;[\;\;{\bf a}^{\mu\nu}(x), {\bf a}^{\lambda\sigma}(y)\;\;]\;\; = 
\frac{1}{(2\pi)^3}\int\frac{d^3p}{2p^o}
[\;\; e^{ip\cdot (x-y)} - e^{-ip\cdot(x- y)}\;\;]\;\;f^{\mu\nu,\lambda\sigma},\label{d}
\end{equation}
where 
\begin{equation}
f^{\mu\nu,\lambda\sigma}=g^{\mu\lambda}\; g^{\nu\sigma} + g^{\nu\lambda}\; 
g^{\mu\sigma} - g^{\mu\nu}\; g^{\lambda\sigma}. 
\end{equation}
The Feynman propagator is
\begin{equation}
\Delta^{\mu\nu,\lambda\sigma}(x -y)=<0|\;T({\bf a}^{\mu\nu}(x)\;\;
{\bf a}^{\lambda\sigma}(y)\;)|0>\;=\;\frac{1}{(2\pi)^4}\int d^4p\;e^{i\;p\cdot(x-y)}\;
\Delta^{\mu\nu,\lambda\sigma}(p),
\end{equation}
where
\[ \Delta^{\mu\nu,\lambda\sigma}(p) =
\frac{1}{2}\;\;f^{\mu\nu,\lambda\sigma}\frac{1}{p^2 - i \epsilon}. \]
This is the propagator of the graviton in the interaction representation.\\

As already noted in this superstring theory there is also a massless vector 
following from the $L_0$ condition acting on $\alpha^{\mu}_{-1}|0,p>$ with Lorentz relation
$p^{\mu}|0,p>$=0, due to the $L_1$ condition. With the help of a time like vector $n^{\mu}$
, we can construct another traceless second rank `Newtonian'(N) tensor
\begin{equation}
{\bf a}^{\dag\mu\nu}_{N,1} =\left (n^{\mu}\alpha^{\nu}_{-1} + n^{\nu}\alpha^{\mu}_{-1} -
g^{\mu\nu} (n\cdot\alpha_{-1}\right )|0,p>_{NS}.
\end{equation}
There are several points which are conflicting. While forming the commutator like
equation(\ref{d}) ,we will arrive at a term with $g^{\mu\nu}g^{\lambda\sigma}$ which 
is already there in the graviton propagator. So there will be over counting. Secondly, 
$n\cdot\alpha_{-1}\; \sim \;\alpha_{-1}^o$ and is excluded in the Fock space, but is 
necessary for vanishing of the trace.\\

We now proceed to construct the propagators, remembering that this is local and 
there is no pole  at $|{\bf p}|^2={p_o}^2$ except in the $g^{\mu\nu}g^{\lambda\sigma}$ term
\begin{equation}
\Delta^{\mu\nu,\lambda\sigma}_{N,1}=\frac{1}{2}\left (\;\frac{f^{\mu\nu\lambda\sigma}_N}
{{|\bf p|}^2} - \frac{ g^{\mu\nu}\;n^{\lambda}\;n^{\sigma} + 
g^{\lambda\sigma}\;n^{\mu}\;n^{\nu}}{{|\bf p|}^2}- \frac{ 
g^{\mu\nu}g^{\lambda\sigma} }{ p^2 }\right ),
\end{equation}
where
\begin{equation}
f^{\mu\nu\lambda\sigma}_N = g^{\mu\lambda}\;n^{\nu\sigma} + g^{\mu\sigma}\;
n^{\nu\lambda} + g^{\nu\lambda}\;n^{\mu\sigma}+  g^{\mu\lambda}\;g^{\nu\sigma}
- g^{\mu\nu}\;n^{\lambda}n^{\sigma}- g^{\lambda\sigma}\;n^{\mu}n^{\nu}.
\end{equation}
We have still the traceless tensor $\Pi^{\mu\nu}(p)$~\cite{we} and the creation 
operator $L_{-1}|o,{\bf p}>$. Since \\$<o,p| L_1\;L_{-1} |o,p>$=2, 
we construct a third traceless tensor,
\begin{equation}
{\bf a}^{\dag\mu\nu}_{N,2}=\frac{1}{\sqrt{2}}\Pi^{\mu\nu}\;L_{-1}\;|o,p>_R.
\end{equation}
The propagator is simply
\begin{equation}
\Delta^{\mu\nu,\lambda\sigma}_{N,2} = \frac{1}{2(p^2-i\epsilon)}\Pi^{\mu\nu}
\Pi^{\lambda\sigma}.
\end{equation}
The gradient terms do not contribute when contracted with conserved energy momentum tensor.
Ignoring the gradient terms 
\[ \Pi^{\mu\nu}\rightarrow g^{\mu\nu} +\frac{p^2}{|{\bf p}|^2}n^{\mu}\;n^{\nu},\]
leading to
\begin{equation}
\Delta^{\mu\nu,\lambda\sigma}_{N,2}=\frac{1}{2(p^2-i\epsilon)}\left ( g^{\mu\nu}g^{\lambda\sigma} +
(g^{\mu\nu}n^{\lambda}n^{\sigma}+n^{\mu}n^{\nu}g^{\lambda\sigma})\frac{p^2}{|{\bf p}|^2}
+n^{\mu}n^{\nu}n^{\lambda}n^{\sigma}\frac{(p^2)^2}{|{\bf p}|^4}\right ).
\end{equation}
The total `Newtonian' graviton propagator is
\begin{equation}
\Delta^{\mu\nu\lambda\sigma}_{N} =\frac{1}{2}\frac{ f^{\mu\nu\lambda\sigma} } 
{ |{\bf p}|^2} +\frac{1}{2}n^{\mu}n^{\nu}n^{\lambda}n^{\sigma}\frac{(p^2)}{|{\bf p}|^4}.
\end{equation}
In all 
\begin{equation}
\Delta^{\mu\nu\lambda\sigma}_{F} = \Delta^{\mu\nu\lambda\sigma}+\Delta^{\mu\nu\lambda\sigma}_{N}.
\end{equation}
In space time, the fourier transformed propagator is
\begin{eqnarray}
\Delta^{\mu\nu\lambda\sigma}_{N}(x-y)
& =&\frac{1}{(2\pi)^4}\int d^4p\;e^{ip\cdot(x-y)}\Delta^{\mu\nu\lambda\sigma}_{N}(p) 
\nonumber\\
 &=&\frac{1}{2}\; \left [ \left ( f^{\mu\nu\lambda\sigma}_N +
n^{\mu}n^{\nu}n^{\lambda}n^{\sigma}\right )\delta(x^o-y^o)\;D({\bf x}-{\bf y}) + 
n^{\mu}n^{\nu}n^{\lambda}n^{\sigma}\ddot{\delta}(x^o-y^o)E({\bf x}-{\bf y})\;\right ],
\end{eqnarray}
where
\begin{equation}
E({\bf x})=\frac{1}{(2\pi)^3}\int d^3q e^{i{\bf q.x}}\frac{1}{|{\bf q}|^4}
=E(o)-\frac{|{\bf x}|}{4\pi},
\end{equation}
and
\begin{equation}
D({\bf x})=\frac{1}{(2\pi)^4}\int d^3q e^{i{\bf q.x}}\frac{1}{|{\bf q}|^2}
=\frac{1}{4\pi |{\bf x}|}.
\end{equation}
Thus we have obtained the quantum field theoretic result of Weinberg~\cite{we} from
this new string theory. $\Delta^{\mu\nu\lambda\sigma}_{N}(x-y)$ is highly divergent
and to cancel this we must add to the free Hamiltonian $H_o$, an interaction
Hamiltonian $H_N'(t)$ as specified below
\begin{eqnarray}
H_N'(t)&=&\frac{1}{2} \int\int d^3x\;d^3y \;\; (\;\; 2\theta^{\mu}_o({\bf x},t)
\theta_{\mu o}({\bf y},t) - \frac{1}{2} \theta^{\mu}_{\mu}({\bf x},t)
\theta_{o o}({\bf y},t)\nonumber\\
&-& \frac{1}{2} \theta_{o o}({\bf x},t)
\theta_{\mu}^{\mu}({\bf y},t)
- \frac{1}{2} \theta_{o o}({\bf x},t)
\theta_{o o}({\bf y},t) \;\;)
D({\bf x}-{\bf y})         
+\frac{1}{2} \int \int  d^3x\;d^3y\;  \theta_{o o}({\bf x},t)
\ddot{\theta}_{o o}({\bf y},t)  E({\bf x}-{\bf y}).
\end{eqnarray}
As Weinberg puts it, `this term  ugly as it seems, is precisely
what is needed to generate Einstein Field Equations when we pass to Heisenberg
representation'. $\theta_{\mu\nu}$ is the symmetric energy momentum tensor with 
vanishing divergence and includes the self energy of the gravitational field, the matter 
field and the Pauli term.
In the Heisenberg representation, the spatial components are defined as
\begin{equation}
a_H^{ij}(x) - \frac{1}{3}\delta^{ij}\delta^{kl}a_H^{kl}(x)= U(x^o)a^{ij}(x) U^{-1}(x^o),
\end{equation}
where
\[ U(x^o) = e^{iH_o t}\;\;\;  and\;\;\;  \partial_{\mu}\partial^{\mu} \; a^{ij}(x)=0,\;\;\;\;\;\; \partial_i a^{ij}(x)
=0.\]
The only nonvanishing commutator is
\begin{equation}
\left [a^{ij}(x),\; \dot{a}^{kl}(y)\right ] = i\;D^{ij,kl}({\bf x-y}).
\end{equation}
To evaluate the D's from the derived propagation function , we introduce a four vector
$\hat{p}^{\mu}=(1, \hat{p})$ and note that
\begin{equation}
\hat{p}^{\mu} = \frac{p^{\mu} + n^{\mu}(|{\bf p}| - p_o)}{|{\bf p}|}.
\end{equation}
The terms containing $p^{\mu}$ when contracted with conserved currents vanish. 
So we have the effective equality
\begin{equation}
\frac{n^{\mu}n^{\nu}}{|{\bf p}|^2} = \frac{1}{p^2} \left [ \hat{q}^{\mu}n^{\nu}+
\hat{q}^{\nu}n^{\mu}-\hat{q}^{\mu}\hat{q}^{\nu} \right ] +\;\;\; gradient \;\;\; terms.
\end{equation}
With the poles at $p^2$, the Green's functions are easily calculated. After some algebra, 
retaining the spatial parts, we get
\begin{eqnarray}
D^{ij,kl}({\bf x-y})& =&\frac{1}{2}\;\;[ \delta^{ik}\delta^{jl}+\delta^{il}\delta^{jk}-\delta^{ij}
\delta^{kl}\;\;]\delta^{(3)}({\bf x-y})\nonumber\\
&+&\frac{1}{2} [ \partial^i\partial^k\delta^{jl} + \partial^i\partial^l\delta^{jk}+ 
\partial^l\partial^k\delta^{il}- \partial^i\partial^j\delta^{kl}- \partial^k
\partial^l\delta^{ij}] D({\bf x-y})\nonumber\\
&+&\frac{1}{2}\partial^i\partial^j\partial^k\partial^l\;E({\bf x-y}).
\end{eqnarray}

The solutions to the Heisenberg field equations are easily worked out and are given by 
Weinberg~\cite{we}. With $\theta_{H,kl}$ as the energy momentum tensor in the Heisenberg 
representation
\begin{equation}
\partial^{\mu}\partial_{\mu}\left [\;\; a_H^{ij}({\bf x},t)-\frac{1}{3}\delta^{ij}\delta_{kl}a_H^{kl}
({\bf x},t)\;\;\right ]
=-\int d^3y\;D^{ij,kl}({\bf x}-{\bf y})\theta_{H,kl}({\bf y},t).
\end{equation}
With Weinberg we also invent the time derivatives from the direct contact term
\begin{eqnarray}
H'_N(t)&=&\frac{1}{2}\int\int \;\;d^3xd^3y\;\;(\;\;  [\;\; 2\;\theta^i_o({\bf x},t)
\theta_{io}({\bf y},t)
-\frac{1}{2}\theta^i_i({\bf x},t)\theta_{oo}({\bf y},t)\nonumber\\
&-&\frac{1}{2}\theta_{oo}({\bf x},t)\theta_{i}^i({\bf y},t)-
\frac{1}{2}\theta_{oo}({\bf x},t)\theta_{oo}({\bf y},t)\;\; ]D({\bf x-y})\nonumber\\
&+&\theta_{oo}({\bf x},t)\ddot{\theta}_{oo}({\bf y},t)E({\bf x-y})\;\;).
\end{eqnarray}
and define
\begin{eqnarray}
&a_H^{io}({\bf x},t)&=\int d^3y\;\theta_H^{io}({\bf y},t)D({\bf x-y});\;\;\;\;
\nabla^2a_H^{io}(x)= -\theta_H^{io}(x),\nonumber\\
&a_{iH}^{i}({\bf x},t)&=\frac{3}{2}\int d^3y\;\theta_H^{oo}({\bf y},t)D({\bf x-y});\;\;\;\;
\nabla^2 a_{iH}^{o}(x)=-\frac{3}{2}\theta_H^{oo}(x),\nonumber\\
&a_H^{oo}({\bf x},t)&=\frac{1}{2}\int d^3y\; [ \theta_H^{oo}({\bf y},t)+
\theta^{i}_{Hi}({\bf y},t) ] D({\bf x-y})
-\frac{1}{2}\int d^3y\ddot{\theta}^{oo}({\bf y},t)E({\bf x-y}),\nonumber\\
&\nabla^2a^{oo}_H(x)& =-\frac{1}{2}\theta^{i}_{Hi}(x)-\frac{1}{2} \theta^{oo}_{H}(x)
+\frac{1}{3}\ddot{a}_{Hi}^i(x) .
\end{eqnarray}
Using tracelessness condition and current conservation condition
\begin{eqnarray}
&\partial_i a_H^{ij}(x)& = \frac{1}{2}\partial^j\theta_{Hi}^i(x),\nonumber\\
&\partial_i a_{Ho}^{i}(x)& = -\frac{2}{3}\partial_o\theta_{Hi}^i(x).
\end{eqnarray}
and after some algebra we arrive at the result obtained by Weinberg
\begin{equation}
R^{\mu\nu}_H(x)= -\theta^{\mu\nu}(x) +\frac{1}{2} g^{\mu\nu}\theta^{\lambda}_{\lambda}(x),
\end{equation}
where
\begin{equation}
R^{\mu\nu}_H(x) =\partial^{\mu}\partial_{\mu} a_H^{\mu\nu}(x)-\partial^{\mu}\partial_{\lambda}a_H^{\lambda\nu}(x)
-\partial^{\nu}\partial_{\lambda}a_H^{\lambda\mu}(x)+
\partial^{\mu}\partial^{\nu}a_{H\lambda}^{\lambda}(x).
\end{equation}
This equation can be put in the Einsteinian form
\begin{equation}
R^{\mu\nu}_H(x)-\frac{1}{2}g^{\mu\nu} R_{H\lambda}^{\lambda} = -\theta^{\mu\nu}_H(x).
\end{equation}

Thus the presence of the tachyons in superstring theory has provided the massless
 states which have led to the construction of graviton and Newtonian graviton and finally
enabled us to deduce the Einstein's field equations following Weinberg. This was first 
deduced by S.N Gupta~\cite{gu}.We have made a direct contact from spin 2 quanta 
string amplitude to the field of the graviton with the help of the tachyonic vacuum. 
To our knowledge, this is the first direct derivation of the Einstein's field equation
from the superstring theory following Weinberg~\cite{we}.
Since superstring theory is renormalisable,we hope that our research will help in 
probing further into the subtelities of Quantum Gravity.

\end{document}